\documentclass{llncs}
\usepackage{graphicx}
\usepackage{amsmath}
\usepackage{amssymb}
\usepackage{amsfonts}
\usepackage[bookmarks]{hyperref}
\usepackage{multirow}
\usepackage{array}
\usepackage{float}
\usepackage{colortbl}
\usepackage{arydshln}
\usepackage{stmaryrd}
\usepackage{pifont}

\makeatletter
\renewcommand\subsubsection{\@startsection{subsubsection}{3}{\z@}%
                       {-18\p@ \@plus -4\p@ \@minus -4\p@}%
                       {0.5em \@plus 0.22em \@minus 0.1em}%
                       {\normalfont\normalsize\bfseries\boldmath}}
\makeatother
\begin{document}

\title{Another Look at Privacy-Preserving Automated Contact Tracing}

\author{Qiang Tang}

\institute{Luxembourg Institute of Science and Technology (LIST)\\
4362, Esch sur Alzette, Luxembourg\\
qiang.tang@list.lu
}

\maketitle



\begin{abstract}
In the current COVID-19 pandemic, manual contact tracing has been proven very helpful to reach close contacts of infected users and slow down virus spreading. To improve its scalability, a number of automated contact tracing (ACT) solutions have proposed and some of them have been deployed. Despite the dedicated efforts, security and privacy issues of these solutions are still open and under intensive debate. In this paper, we examine the ACT concept from a broader perspective, by focusing on not only security and privacy issues but also functional issues such as interface, usability and coverage. We first elaborate on these issues and particularly point out the inevitable privacy leakages in existing BLE-based ACT solutions. Then, we propose a venue-based ACT concept, which only monitors users' contacting history in virus-spreading-prone venues and is able to incorporate different location tracking technologies such as BLE and WIFI. Finally, we instantiate the venue-based ACT concept and show that our instantiation can mitigate most of the issues we have identified in our analysis.
\end{abstract}

\section{Introduction}
\label{sec:intro}

In the public health domain, \emph{contact tracing} refers to the process of identifying users who have come into close contact with an infected patient and subsequently collecting further information about them. By tracing the contacts, testing them for infection, treating the infected and tracing their contacts in turn, the ultimate infection rate of the disease in the population can be significantly reduced. Today, contact tracing has been widely performed for diseases like sexually transmitted infections (e.g. HIV) and virus infections (e.g. SARS-CoV). It has been proven effective in combating contiguous diseases because it can at least (1) interrupt ongoing transmission and reduce spread, alert contacts about the possibility of infection and offer preventive counseling or prophylactic care, and (2) allow the medical professionals to learn about the epidemiology in a particular population. Although various digital technologies might be leveraged, the contact tracing task is often carried out manually by trained human experts through calling/meeting the infected patients to obtain necessary information of their contacts and then reaching out to these contacts.

Starting from the end of 2019, the Coronavirus disease (referred to as COVID-19), caused by the SARS-CoV-2 virus, has drastically disrupted the society and no effective medical solution has been found until today. The SARS-CoV-2 virus is highly contagious and its transmission paths are very versatile. Medical experts believe that it can spread via respiratory transmission, contact transmission, and aerosol transmission as well. This essentially means that a user might be at risk if he stays close to a patient for a certain amount of time, toughs a surface tainted by a patient, or enters into a closed space where a patient has just stayed. From the beginning of the pandemic, manual contact tracing has been carried out in most countries to trace at-risk users and help slow down virus spreading. Indeed, it has been proven useful, and many regional/national governments have expanded their contact tracing teams over the time. However, due to the sophisticated nature of SARS-CoV-2 virus, people realised that manual contact tracing faces an efficiency (or, scalability) problem. To this end, Ferretti et al. \cite{Ferretti20} investigated the key parameters of epidemic spread and concluded that viral spread is too fast to be contained by manual contact tracing but could be controlled if this process was faster, more efficient and happened at scale. This essentially necessitates \emph{automated contact tracing} (ACT) solutions, which can automatically record mutual contacts without human intervention. Later on, the theoretical result of Ferretti et al. was validated by Tian et al.  in their empirical analysis of the control measures from China \cite{Tian20}.

Note that our notion of ACT solutions refers to those that rely on either proximity or Geo-location data to classify contacts. They belong to a larger category of digital contract solutions which broadly rely on digital technologies to facilitate contact tracing, as surveyed by Redmiles in \cite{Redmiles2020}.

\subsection{Automated Contract Tracing}
\label{subsec:arch}

The core of ACT solutions is location tracking technologies, which can automatically determine the physical distance between users without the need of any human intervention. Among all (see a detailed survey in \cite{Nguyen2020}), Geo-location technologies (e.g. those based on GPS, WIFI, Telcom Cell Towers) and Bluetooth Low Energy (BLE) technologies are the popular choices, and BLE is the mostly used one due to its technical easiness and privacy friendliness. From a user's perspective, most ACT solutions are in the form of apps although they create sophisticated cyber-physical-social ecosystems. Usually, these apps can be installed on smartphones as well as other compatible electronic devices. Throughout the paper, we will use ``apps" and ``ACT solutions" interchangeably, unless a distinction is necessary.

\begin{figure}[h]
\centering
\includegraphics[scale=0.46]{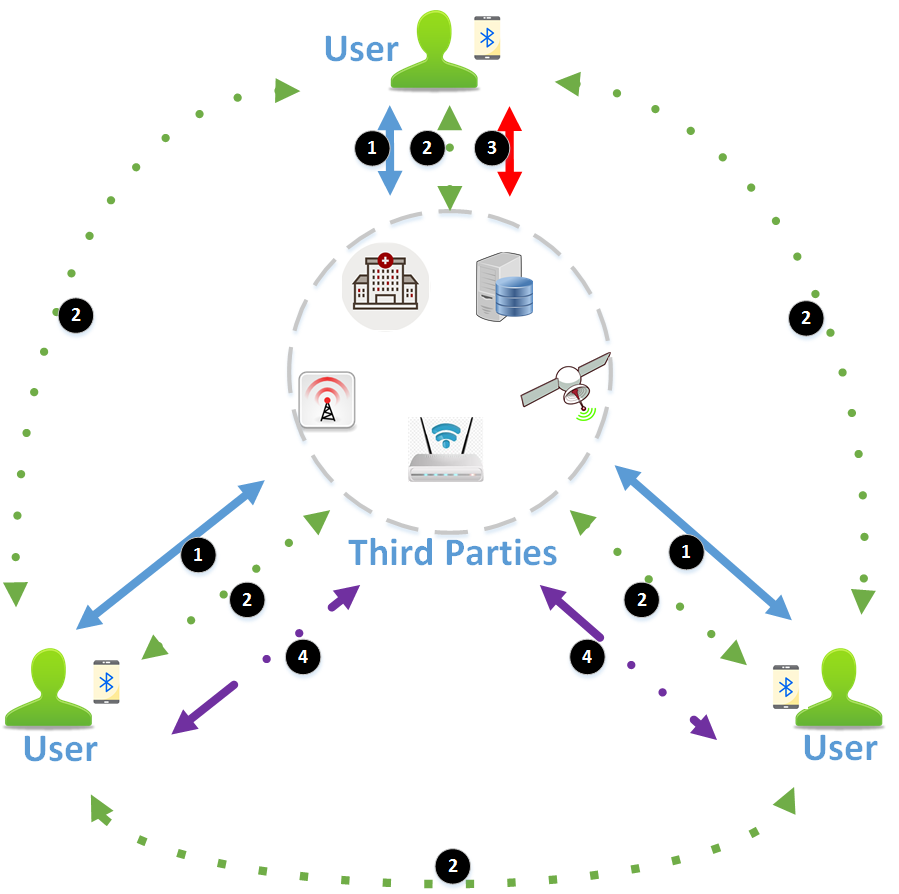}
\caption{System Architecture}
\label{architecture}
\end{figure}

Referring to \cite{Tang2020}, an ACT solution generally involves two categories of players. One category comprises all relevant third parties, which consist of health authorities, medical personnel and potentially others.  The other category comprises individual users, who are supposed to install an app and interact with the third parties.  As illustrated in Fig. \ref{architecture}, the workflow of an ACT solution can be logically divided into four phases, namely (\emph{initialisation}, \emph{sensing}, \emph{reporting}, \emph{tracing}).

In the \emph{initialisation} phase, individual users and relevant third parties need to set up the system to enable the operations in other phases. For example, every individual user might be required to have a smart phone and download an app from a third party. In addition, cryptographic credentials may need to be generated and distributed. In the \emph{sensing} phase, individual users will record their own location trails and also collect location data from their close contacts. In the \emph{reporting} phase, an individual user firstly makes a test. If the result is positive, she needs to collaborate with some third parties (e.g. health authority) to make her relevant location data available for the further uses. In the \emph{tracing} phase, relevant third parties could collect and aggregate the location data from infected individuals for any possible legitimate purposes. For instance, a third party can evaluate the infection risks of the close contacts and communicate with them accordingly, or let them evaluate the infection risks on their own. We note that this workflow aligns with the standard testing-tracing framework of manual contact tracing.

\subsection{The Current Situation}

In practice, an ACT solution can involve a large number of players. Besides users, health authorities and medical personnel, it also engages app developers, device manufacturers, operating system providers (e.g. when Google-Apple API is used), cloud storage service providers, regulators, and so on. This complicates the governance aspect of such solutions and makes it extremely difficult to strike a balance among all desired properties (e.g. functional properties described in Section \ref{sec:funchallenges} and security and privacy properties described in Section \ref{sec:spchallenges}). The security and privacy issues associated with existing ACT solutions have already been discussed in many places, e.g. \cite{Gvili2020,Krishnan2020,Vaudenay20202,Vaudenay20203}. Particularly, Vaudenay and Vuagnoux \cite{Vaudenay20203} performed a thorough analysis against the Swiss solution \emph{SwissCovid} and identified a large number of issues w.r.t. security and privacy.

In the real-world deployment, more than 30 apps have been proposed according to MIT Technology Review's contact tracing tracker project\footnote{\url{https://www.technologyreview.com/2020/05/07/1000961/launching-mittr-covid-tracing-tracker/}}.
Among the pioneers, Singapore's \emph{TraceTogether} has been installed by around 30\% of the country's residents since its release in mid March 2020, and Iceland's \emph{Rakning C-19} has remarkably been installed by 40\% of the country's residents since its release in early April 2020\footnote{\url{https://www.technologyreview.com/2020/05/11/1001541/iceland-rakning-c19-covid-contact-tracing/}}. Unfortunately, up to now, it is widely perceived that these deployed solutions have not helped much in tracing at-risk users and flattening the COVID-19 curve\footnote{\url{https://www.technologyreview.com/2020/05/11/1001541/iceland-rakning-c19-covid-contact-tracing/}}. Even with a low adoption rate, an ACT solution could still help to some extent as shown by Abueg et al. \cite{abueg2020}, but Hinch et al. \cite{Hinch2020} indicate that a higher adoption rate is required to make an ACT solution significantly useful in practice.  It is worth mentioning that some countries such as Ireland and Germany think their solutions have been a success\footnote{\url{https://www.technologyreview.com/2020/08/10/1006174/covid-contract-tracing-app-germany-ireland-success/}}, although no further details have been made public.

Some researchers have attributed the low adoption rates of the ACT solutions to the underlying security and privacy concerns. It is clearly true that such concerns have prevented many users from downloading and installing the apps. However, we would like to argue that the range of potential issues facing such apps has been underestimated. This is why, despite all the efforts and initiatives, today, it remains an open question to properly position ACT inside the toolbox of fighting the COVID-19 pandemic and come up with a realistic solution.

\subsection{Our Contribution}

To facilitate our analysis, we first recap two BLE-based solutions, namely Singapore's \emph{TraceTogether} and the DP-3T solution by Troncoso et al. \cite{dp3t2020} in Section \ref{sec:review}. Moreover, we categorize and survey some recent ACT solutions with a focus on BLE-based ones. In Section \ref{sec:funchallenges}, we investigate three functional issues facing existing ACT solutions. The first issue is lacking necessary interfaces to other types of solutions. On one hand, it is a useful feature to prevent privacy leakage, but on the other hand it downgrades an ACT solution's efficacy in practice. The efficacy degradation is also reflected in a usability issue, namely lacking explanation and intervention of human experts. We finally examine the coverage issue from the perspective of both population coverage and transmission path coverage. In Section \ref{sec:spchallenges}, we investigate the security and privacy issues which have already been studied in the literature. We enumerate a number of security concerns and highlight the consequences due to the lacking authentication or binding.  Regarding privacy, we highlight three observations: \emph{persistent surveillance} which leaves a 24/7 attack surface for attackers, \emph{ephemeral linkage} which allows attackers to track users in many subtle scenarios, and \emph{violation of data minimisation principle} which leaks unnecessary information about the infected users. These concerns root in the design pattern of existing ACT solutions and are very difficult to be addressed in the current paradigm.

Based on our analysis, in Section \ref{sec:concept} we propose a venue-based ACT concept with two new features. One  is \emph{geo-selective tracing} which means that users' contact history is only monitored in virus-spreading-prone venues. This feature reduces the attack surfaces for both security and privacy without affecting the efficacy. The other is \emph{accountable mediating}, which brings venues as a new type of player into the loop. Accountable venues can improve security and privacy protection by deploying cybersecurity tools on their physical premise. Moreover, under users' consent, venues can take advantage of the collected data to address the functional issues, e.g. expanding coverage by integrating manual contact tracing procedures and running privacy-preserving data analysis protocols to gain more insights into the virus spreading patterns. In Section \ref{sec:instantiation} we instantiate the venue-based ACT concept by describing a new workflow and three sub-protocols for venues who support BLE technologies. We further show that our instantiation mitigates most of the issues we identified against existing ACT solutions. In Section \ref{sec:con} we conclude the paper.

\section{Brief Survey of Existing Solutions}
\label{sec:review}

In this section, we first review two BLE-based solutions which illustrate the centralised and decentralised design philosophies respectively. Note that the DP-3T solution by Troncoso et al. \cite{dp3t2020} has been the basis of many other solutions, e.g. the Exposure Notification API from Google and Apple\footnote{\url{https://covid19.apple.com/contacttracing}}. We then review some other interesting solutions, and refer the readers to dedicated surveys such as  \cite{Ahmed2020,Reichert2020} for more information about the subject.

\subsection{Centralised vs. Decentralised Solutions}
\label{subsec:examples}

In some centralised solutions, the health authority is capable of identifying the at-risk users and is authorised to contact and instruct them for further actions. While, in other centralised solutions, the authority can push information to the at-risk users even if it cannot straightforwardly identify the users. In contrast, in decentralised solutions, the authority is not supposed to maintain any contact details of the users, and is only authorised to publish certain collected information from the infected patients. A user's app is responsible for downloading the published information and computing the infection risk score.

The \emph{TraceTogether} protocol from Singapore is a centralised solution, by assuming that Ministry of Health (MoH) of the Singapore government is fully trusted. Let assume there are ($1 \leq i \leq N$) users, the protocol is elaborated below.

\begin{itemize}
\item In the \emph{initialisation} phase, a user $i$ downloads the \emph{TraceTogether} app and installs it on her smartphone. The app sends the phone number $NUM_i$ to MoH and receives a pseudonym $ID_i$. MoH stores the ($NUM_i, ID_i$) pair in its database. MoH generates a secret key $K$ and selects an encryption algorithm $\mathsf{Enc}$. At the beginning of the app launch, MoH decides some time intervals $[t_0, t_1, \cdots]$, which will end when the pandemic is over. For the user $i$, MoH pushes $TID_{i,x} = \mathsf{Enc}(ID_i, t_x; K)$ to user $i$'s app at the beginning of $t_x$, for $x \geq 0$.


\item In the \emph{sensing} phase, user $i$ broadcasts $TID_{i,x}$ at the time interval $[t_x, t_{x+1})$ for every $x \geq 0$. For example, when user $i$ and user $j$ come into a range of Bluetooth communication at the interval $[t_x, t_{x+1})$, then they will exchange $TID_{i,x}$ and $TID_{j,x}$. They will store a $(TID_{i,x}, TID_{j,x}, Sigstren)$ locally in their smartphones, respectively. The parameter $Sigstren$ indicates the Bluetooth signal strength between their devices.


\item In the \emph{reporting} phase, suppose that user $i$ has been tested positive for COVID-19, then she is obliged to share with MoH the locally-stored pairs $(TID_{i,x}, TID_{j,x}, Sigstren)$ for all relevant $j$ and $x$.


\item In the \emph{tracing} phase, after receiving the pairs from user $i$, MoH decrypts every $TID_{j,x}$ and obtains $ID_j$. Based on $ID_j$, MoH can looks up $NUM_j$ and then contact user $j$ for further instructions.

\end{itemize}

Note that even if MoH is assumed fully trusted, the users are not required to share everything with MoH if they have not been in close contact with any COVID-19 patient.

Next, we first recap the low-cost version of the DP-3T solution, and then briefly mention its two enhanced variants. The solution assumes user $i$ $(1 \leq i \leq N)$, a back-end server, and a Health Authority (HA). The back-end server acts as a communication platform to facilitate the matching messages among the users. Let $\mathsf{H}$, $\mathsf{PRG}$ and $\mathsf{PRF}$ denote a cryptographic hash function, a pseudorandom number generator and a pseudorandom function, respectively.

\begin{itemize}
\item In the \emph{initialisation} phase, user $i$ generates a random initial daily key $SK_{i,0}$, and computes the following-up daily keys based on a chain of hashes: i.e. the key for day 1 is $SK_{i,1}=\mathsf{H}(SK_{i,0})$ and the key for day $x$ is $SK_{i,x}=\mathsf{H}(SK_{i,x-1})$. Suppose $n$ ephemeral identifiers are required in one day, then the identifiers for user $i$ on the day $x$ are generated as follows:
    \begin{displaymath}
    EphID_{i, x, 1}||\cdots||EphID_{i, x, n}= \mathsf{PRG}(\mathsf{PRF}(SK_{i,x}, ``broadcast key"))
    \end{displaymath}


\item In the \emph{sensing} phase, on the day $x$, user $i$ broadcasts the ephemeral identifiers $\{EphID_{i, x, 1}, \cdots, EphID_{i, x, n}\}$ in a random order. At the same time, her app stores the received ephemeral identifiers together with the corresponding proximity (based on signal strength), duration, and other auxiliary data, and a coarse time indication (e.g., ``The morning of April 2”).


\item In the \emph{reporting} phase, if user $i$ has been tested positive for COVID-19, then HA will instruct her to send $SK_{i,x}$ to the backend server, where $x$ is the first day that user $i$ becomes infectious. After sending the $SK_{i,x}$ to the backend server, user $i$ chooses a new daily key $SK_{i,y}$ depending on the day when this event occurs.


\item In the \emph{tracing} phase, periodically, the backend server broadcasts $SK_{i,x}$ after user $i$ has been confirmed with the infection. On receiving $SK_{i,x}$, user $j$ can recompute the ephemeral identifiers for day $x$ as follows

    \begin{displaymath}
    \mathsf{PRG}(\mathsf{PRF}(SK_{i,x}, ``broadcast key")).
    \end{displaymath}
    Similarly, user $j$ can compute the identifiers for day $x+1$ and so on. With the ephemeral identifiers, user $j$ can check whether any of the computed identifiers appears in her local storage. Based on the associated information, namely ``proximity, duration, and other auxiliary data, and a coarse time indication", user $j$ can act accordingly.

\end{itemize}

To further enhance its privacy guarantee, two variants have been proposed in \cite{dp3t2020}. In one variant, selected ephemeral identifiers of infectious user $i$ are hashed into a Cuckoo filter, which will then be shared with the public. The other variant lies in the middle of the low-cost version and the first enhancement to achieve a trade-off between privacy protection and complexity.

\subsection{More Related Work}
\label{subsec:related}

Against the popular argument that decentralised solutions offer more privacy protection than the centralised ones, the authors in \cite{Fraunhofer2020,Vaudenay20202} carried out thorough analysis and showed that decentralised solutions are more vulnerable to \emph{targeted surveillance} while the centralised ones are more vulnerable to \emph{mass surveillance}. It is clearly unfair to claim that one category is superior to the other. Nevertheless, most research work has focused on designing decentralised solutions which exhibit two different design patterns.

One pattern is \emph{upload-what-you-sent}.  In this case, an infected patient Alice uploads the messages she has sent (or, equivalently the private keys used to generate these messages as in the DP-3T solution) to an authority, which aggregates  the messages from all infected patients and shares them with the public. Based on how many of these messages have been collected, Bob's app can calculate the infection risk for Bob. Besides DP-3T, similar solutions like PACT-EAST \cite{pacteast2020}, PACT-WEST \cite{pactwest2020}, Canetti-Trachtenberg-Varia \cite{canetti2020} fall into this category. Avitabile \cite{Avitabile2020} proposed a solution based on public bulletin board (e.g. Blockchain), blind signature and Diffie-Hellman key exchange protocol. Each user anonymously posts her ephemeral Diffie-Hellman public keys (i.e. one for each time epoc)  to the bulletin board and memorizes their addresses. At each time epoc, a user Alice's app broadcasts her corresponding public key address, and collects those from the encounters. If Alice is tested positive, based on the collected public key addresses, she fetches the corresponding ephemeral Diffie-Hellman public keys from the bulletin board and computes the Diffie-Hellman shared keys.  Alice then asks the medical lab (which certifies her infection) to blindly sign these Diffie-Hellman shared keys and uploads them to the bulletin board. Later on, a user Bob can compute Diffie-Hellman shared keys based on the collected public key addresses in his local storage, and compare them with those on the bulletin board. If there is a match, then he might be at risk of infection. Trieu et al. \cite{Trieu2020} proposed a solution based on private set interaction (PSI) protocols.  A user Bob (who has collected pseudorandom identifiers from his encounters) runs a PSI protocol with a server (who collects all the  pseudorandom identifiers emitted by infected users) to evaluate his risk of infection based on the size of the intersection of their identifier sets. Pinkas and Roneny \cite{Pinkas2020} proposed a solution with a number of new features. For instance, it offers a tradeoff between privacy and explanability for the at-risk users, and it also provides mechanisms based on coarse geo-location information to guarantee authenticity and prevent replay attacks. In addition, the solution also allows an at-risk user to prove to a server the fact of his exposure.

The other design pattern is \emph{upload-what-you-heard}. In this case, an infected patient Alice uploads the messages she has received to the authority, which aggregates  the messages from all infected patients and share them with the public. Based on how many of these messages have been generated by itself, Bob's app can calculate the infection risk for Bob. Beskorovajnov et al. \cite{Beskorovajnov2020} proposed a solution that aims at enhancing the privacy protection of infected users and facilitating an at-risk user to prove the fact, by introducing multiple backend servers. Different from other solutions, an app generates a pair of pseudorandom identifiers for each time epoc and anonymously shares them with a submission server. At each epoc, the app broadcasts the first element of the relevant pair and removes it from local storage while keeping the second element. At the same time, the app also collects identifiers emitted by the encountered peers. If a user Alice is tested positive, she uploads the relevant collected identifiers to a matching server. Then, for each collected identifier (which is the first element of a pair from one user) received by the matching server, a notification server tries to find out the corresponding second element and publishes it.  Finally, another user Bob can download the published second elements from the notification server and match with all the second elements of his own pairs (where the first elements of those pairs have been deleted). Liu et al. \cite{Liu2020} proposed a solution based on cryptographic primitives including group signature and zero-knowledge proofs. If Alice is tested positive, she gives medical personnel a pseudo-randomized public key for each of her close contact together with a zero-knowledge proof to prove the contact fact. The medical personnel stores the pseudo-randomized public keys on a public bulletin board, which allows any user Bob to check whether one pseudo-randomized public key belong to him. Canetti et al. \cite{canetti2020-1} proposed two solutions, named \emph{ReBabbler} and \emph{CleverParrot} respectively. \emph{ReBabbler} adopts an \emph{upload-what-you-sent} approach and  \emph{CleverParrot} adopts an \emph{upload-what-you-heard} approach. Both solutions construct their identifiers based on local time stamps and private seeds, and the \emph{CleverParrot} solution works only if the users have strong synchronization in their clocks.

From the perspective of infected patients, solutions with a \emph{upload-what-you-sent} pattern pose higher privacy risk because messages emitted by these patients are directly shared with the public. In contrast, solutions with a \emph{upload-what-you-heard} pattern offer better privacy protection. The downside is that these solutions are substantially more complex w.r.t. system architecture and/or required computation than the former category. Note that there are hybrid solutions, which combine the features of centralised and decentralised ones. For instance, Castelluccia et al. \cite{Castelluccia2020} proposed a solution based on Diffie-Hellman key exchange, where the ephemeral messages from Alice and Bob are in the form of $g^a$ and $g^b$ respectively. If Alice is tested positive, then she will submit $g^{ab}$ to the back-end server, which will evaluate Bob's infection risk if he also submits $g^{ab}$.

\section{Functional Issues of ACT Solutions}
\label{sec:funchallenges}

To set up the scene for our discussion, we make a \emph{democratic assumption} that every legitimate user is granted the right to voluntarily adopt a solution (i.e. install and enable a tracing app) and freely decide what to do after receiving an infection risk score.

\subsection{The Interface \& Usability Issue}
\label{subsec:interface}

Existing ACT solutions unanimously aim at evaluating the infection risk of their users and alerting them if necessary. Putting aside the diverse underlying technologies and risk calculation formulas, the main difference among them is how alert is delivered to an at-risk user. Recall from Section \ref{sec:review}, the authority will inform the at-risk users in a centralised solution, while a user evaluates his risk locally in his app with a decentralised solution. To prevent privacy leakages, it has been advocated that information collected by an ACT solution should not be used for any secondary purpose. There is also a sunset rule which requires the installed app to be automatically deleted after a certain period of time.

Minimizing the interface to the rest of the world is important to safeguard users' privacy for an ACT solution. Unfortunately, this raises some usability concern, which has already been reflected by the inefficacy of existing deployments. We investigate the concern from both the human and technical aspects, and highlight the importance of getting human tracing experts into the loop.

\begin{itemize}

\item After receiving numerical infection risk scores, users may behave very differently. Some users might choose to ignore the (very high) risk scores. Their ignorance can be due to the fact that they do not have any symptom or they do not seriously care about their health. In contrast, other users might become very nervous even the received scores are moderate or minor. For these users, anxiety will be easily aggravated when they are accidentally experiencing certain symptoms (e.g. fever and coughing) as a result of other diseases such as a seasonal flu. In both cases, it is crucial that human tracing experts can step in and communicate with the users.
\begin{itemize}
\item W.r.t. a user from the first group, a human expert can help her confirm the infection risk and possibly persuade her to perform a test. Moreover, the human expert can try to identify the at-risk user's close contacts as well as other valuable information such as the visited places and participated events. With such information, the human expert can continue to communicate with these identified contacts to figure out their infection risks, and at the same time alert the authority to proactively take other necessary measures (e.g. sanitizing some facilities).

\item W.r.t. a user from the second group, a human expert can help evaluate the risk of infection, based on professional knowledge and the user's current symptoms. Through a relaxed conversation, the human expert can help the user eliminate the unnecessary anxiety and provide other useful advice.
\end{itemize}

\item Popular location tracking technologies such as GPS and BLE have their inherent limitations in accurately measuring the physical distances of mobile devices. For example, GPS cannot measure the vertical distance of objects inside a building and BLE can be easily influenced by the environment (e.g. whether users have put their smart phones in their pockets or not \cite{Leith2020}). Inaccurate distance measurements will directly lead to two risks in classification: false positive (a user is classified as at-risk while he is not) and false negative (a user is not classified as at-risk while he is). These risks not only reduce the usefulness of the underlying solution but also cause additional harm to the users and the society (e.g. panic). In order to mitigate the consequences of false positive and false negative results, it is crucial to introduce human tracing experts into the loop as well.
\begin{itemize}
\item Regarding false positives, we can expect a human expert to re-evaluate  infection risk scores by taking into account additional factors. For instance, a user has been to a hospital and drawn contacts with infected patients, but all involved have been in protected mode with face masks and so on. In this case, the infection risk can be regarded as very low, regardless the numerical score based on physical distance measurements.
\item Regarding false negatives, a user may suspect himself being infected even if he only receives a minor score. In this case, we can expect a human expert to re-evaluate the infection risk based on more information from the user. For instance, the user may have been to a party which has already infected patients reported. In this case, the user is likely at risk and should perform a test.
\end{itemize}
\end{itemize}

It is worth noting that false positive and false negative results can also come from active attackers. For instance, a malicious user can create a false positive result (e.g. by putting his smart phone close to that of an infected patient) so that he can stay home for two weeks without working. To resolve such issues, human tracing experts could also play an important role.

In order to empower the users, many ACT solutions have tried to put them into finer control of their location data, e.g. deciding which part of their location data will be shared with the health authority. We want to emphasize that such design decisions may lead to additional false positive and false negative risks, and subsequently downgrade the solutions' efficacy. For example, if infected users only share a small part of their location data due to privacy concerns, many of their contacts will receive false negative alerts. On the other hand, a malicious user may have the opportunity to manipulate his location data to generate false positives for some target contacts.

\subsection{The Coverage Issue}
\label{subsec:coverage}

It is widely known that, when exposed to the SARS-CoV-2 virus (and other viruses as well), the infection risk and the consequences after infection vary a lot for different user groups. The most vulnerable group of people are those who have some base diseases (e.g. diabetes and cancers) and the elderly whose immune system is too fragile to defend against the virus. Therefore, the infection risk of this group should be measured more closely and medical intervention needs to be applied in a more timely manner. However, contrary to this natural requirement, it is estimated that a large proportion of this group do not even have a smart phone and many cannot manage an ACT app. We refer this to be the \emph{population coverage} problem, i.e. the more vulnerable users are less likely to adopt an ACT solution.

Existing solutions also face a \emph{transmission path coverage} problem, because their infection risk calculation formulas only target the respiratory transmission path (i.e. a user is considered at risk if he stays with an infected patient within 2 meters for more than 15 minutes). Given the current risk definitions, \emph{contact and aerosol transmission paths} can not be effectively covered. This results in significant false negatives. For example, an infected patient Alice may cough a lot on a bus and another user Bob sits very close to her for a few minutes. In this case, Bob can be at high risk because he may have contracted the virus by touching contaminated objects. As another example, a patient Alice has stayed in a closed meeting room for 2 hours, during which she coughed and sneezed several times, and afterwards Bob comes in and has a long meeting. In this case, even if Alice and Bob have not stayed together whatsoever, Bob can still be at high risk of contracting the virus.


\section{Security and Privacy Issues of ACT Solutions}
\label{sec:spchallenges}

In Section \ref{sec:funchallenges}, we point out how existing ACT solutions fall short in their functionalities. In this section, we will examine their security and privacy issues and highlight the open challenges.

\subsection{The Security Challenges}
\label{subsec:security}

As shown by Vaudenay and Vuagnoux in analysing the \emph{SwissCovid} solution \cite{Vaudenay20203}, an ACT solution may involve a large number of players. Besides users, health authority and medical personnel, app developers, device manufacturers, operating system providers, cloud storage service providers and regulators are also involved and have their respective impacts on security and privacy. Regarding most solutions including those mentioned in Section \ref{sec:review}, their authors have not described the system and security models with the necessary granularity to cover all these players. As such, it remains an open question what are the precise security (and privacy) guarantees these solutions can offer.

In more detail, we would like to emphasize on the following security issues. Considering the sophisticated nature of ACT, many other issues exist.

\begin{itemize}

\item Authenticity or binding is usually missing between the users, smart phones, apps, and the location data. Malicious users can leverage this fact to carry out fraudulent activities, e.g. a user may use another user's smart phone and infection risk score to get a priority in testing. The lacking of authentication and subsequently trustworthiness of data from the ACT solution makes it impossible for the health authority to provide any further service to the users in an accountable and fair manner.

\item In most solutions, particularly the BLE-based ones, there is neither authentication nor integrity protection for the exchanged messages. This leads to straightforward replay and relay attacks, see e.g. \cite{Vaudenay2020,Vaudenay20202}, in which an attacker can either replay messages directly or relay the messages to be broadcast in different geographic locations. In addition, for transparency reasons, it has been recommended that any ACT protocol and its implementation should be made public so that they can be scrutinized by security experts. This makes it effortless for an attacker to produce fake apps. Leveraging these vulnerabilities, a powerful attacker can launch distributed denial of service (DDoS) attacks, in which it can generate false positives among a large number of users and cause panics among the society. Note that it is difficult to attribute the occurred attacks to the attacker due to the lack of authentication at various levels.

\item Smart phones can be stolen or lost, so that data might be recovered. In reality, smart phones and operating systems have their own security issues, which may facilitate attackers to remotely exploit an ACT app. Similarly, malicious users or coerced benign users might abuse their access control rights to manipulate their data and disrupt the normal operations.

\item Referring to the system specification from Section \ref{subsec:arch},  an infected user needs to share information about her contacts with some third parties (e.g. health authority) in the \emph{reporting} phase, making it possible to alert the at-risk users. In this process, it is important to guarantee that the shared information is faithfully belonging to the infected user. A number of dishonest behaviours could occur in this process. For example, the infected user can provide false contact information, and a malicious user can impersonate the infected user to upload false information. Security issues in this aspect are often beyond the security model of an existing solution, and left to be addressed in the development phase.
\end{itemize}

It could be argued that we have readily-available mechanisms to solve some of the aforementioned security issues (e.g. authentication). However, for ACT solutions, there is an inherent dilemma between system/information security and user privacy. Today, there is not even a consensus on an acceptable trade-off between these requirements.

\subsection{The Privacy Challenges}
\label{subsec:privacy}

The promotion of ACT solutions has triggered tremendous privacy concerns in the society. One concern is \emph{mass surveillance} (i.e. big brother problem), through which a powerful attacker learns some private information about an entire or a substantial fraction of the population. For instance, centralised solutions have been criticized because they can potentially allow the health authorities to learn the social graph of the involved users. Another concern is \emph{targeted surveillance} (i.e. small brother problem), through which an (either ordinary or powerful) attacker learns some private information about some particular users. For example, an attacker can try to learn whether or not a celebrity has been infected by the virus. For BLE-based solutions, a curious user can easily determine who around him has installed a tracing app. Whether or not installing the tracing app might be a voluntary option in general, but disclosing this fact might be a concern for certain users. For instance, if the fact that a governmental official has chosen not to install the app is publicized, then this official might face accountability criticisms from the citizens.

Orthogonal to these two, there is a concern of \emph{persistent surveillance}, which means that the tracing app on a user's smart phone will remain active anytime and anywhere unless the user switches it off manually. Clearly, such surveillance is unnecessary in many circumstances, e.g. during sleep and in well protected areas. Furthermore, it may lead to new privacy leakages. For instance, an attacker may perform precisely targeted attacks, e.g. spying around a celebrity's home or, inferring relationship between users by correlating messages emitted by their smart phones.

Specific for BLE-based solutions, the app installed on a smart phone needs to repeatedly broadcast the same pseudorandom identifier in every time epoc, often a few minutes. This leads to an \emph{ephemeral linkage} issue: two identical pseudorandom identifiers will imply the broadcasters are the same app (assuming no attacker is replaying the messages). This kind of linkage may lead to serious tracking concerns. For instance, an employee Alice can leverage it to easily determine with whom his colleague Bob is having a private meeting in his office, without the need to physically spying around Bob's office. Note that the \emph{ephemeral linkage} issue will be amplified if the attacker is able to collect data in adjacent geographic locations. Another concern is that most BLE-based solutions violate the \emph{data minimisation} principle. Using the DP-3T from Section \ref{subsec:examples} as an example, suppose a user Alice and an infected patient Bob encounter each other on the street, and Alice's smart phone receives one ephemeral identifier from Bob. If the encounter lasts only 10 seconds, then Alice will not have any infection risk so that she should not learn anything about the status of Bob. In reality, Alice can choose to keep the identifier to match it with the identifiers from the backend server to find out whether Bob has been infected. This should be regarded as an unnecessary privacy leakage from Bob's perspective. We note that although an official solution might not allow a user to keep the ephemeral identifier \cite{Vaudenay20203}, Alice can still does it with a homemade app as the ACT protocol is often detailed in public \cite{Baumgartner2020}.

To sum up, some privacy risks are due to the lack of security protections (e.g. authentication), while other risks are inevitably implied by the functionality (e.g. mass surveillance in centralised solutions and targeted surveillance in decentralised solutions). We want to emphasize again that the relationship between privacy and security is very subtle. One one hand, imposing security protections (e.g. asking strong authentication for the user and his app) might result in privacy risks. On the other hand, the lack of security protection facilitates attackers to inject and collect location data from (targeted) users and infer private information afterwards.

\section{New Venue-based ACT Concept}
\label{sec:concept}

As shown in Section \ref{sec:funchallenges} and Section \ref{sec:spchallenges}, existing ACT solutions face not only functional but also security and privacy issues. Due to the sophisticated nature of the application environment, it is very difficult to find universal tradeoffs among all the diverse requirements. Prudent engineering based on state-of-the-art cryptographic primitives might enable us to achieve some specific security and privacy properties, however, it does not seem promising to help us simultaneously address the identified issues. Particularly, some (privacy) issues are inherently implied by the functional design. To avoid these inherent systemic drawbacks, we propose a venue-based ACT concept with two advanced features, namely \emph{Geo-selective Tracing} and \emph{Accountable Mediating}.

\subsection{Geo-selective Tracing}
\label{subsec:trace}

It is natural to assume that virus infection among users mostly occurs in crowded venues where they get a higher chance to engage in long-lasting close contact either directly or indirectly. Some example venues include offices, restaurants, churches, shopping malls, transportation facilities such as trains and buses, sports facilities such as gyms and swimming pools, and schools. This assumption has been practically validated by the existing statistics from various sources as well as the frequently reported cluster infection incidents. Therefore, in order to capture most at-risk users without increasing false negatives, it suffices to deploy a venue-based ACT that traces at-risk users in geographically selected venues. This implies that an ACT solution only needs to be active in selected venues instead of everywhere.

\begin{figure}[h!]
\centering
\includegraphics[scale=0.36]{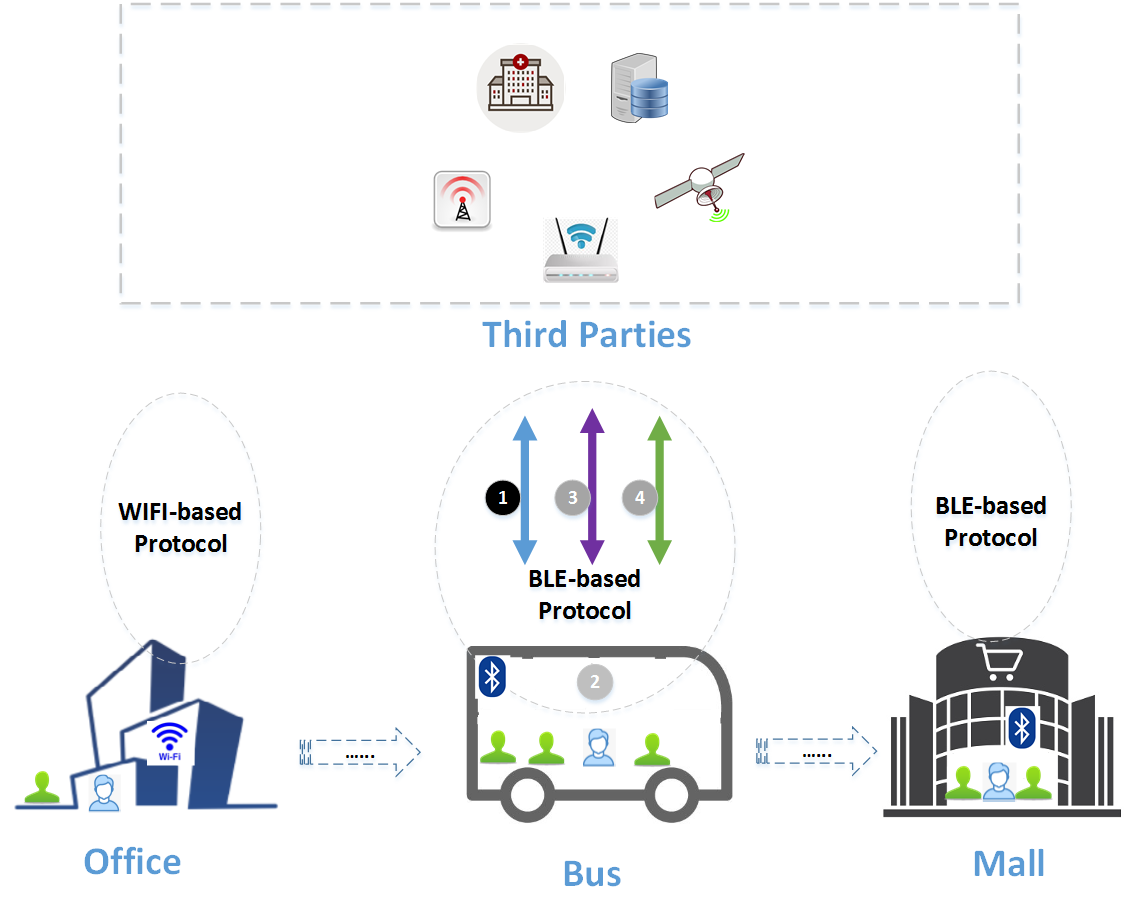}
\caption{New Solution Architecture}
\label{newarchitecture}
\end{figure}

At a very high level, a venue-based ACT solution will possess an architecture shown in Fig. \ref{newarchitecture}.  Functionally, it can be considered as a composition of standard ACT protocol described in Fig. \ref{architecture}. Since contact tracing is essentially carried out independently in different venues, every venue can in principle choose its own location tracking technologies (e.g. WIFI or BLE) and infection risk evaluation formulas, according to the environment characteristics (e.g. indoor or outdoor) and existing facilities (e.g. isolated rooms or open offices). As an example, employees may all use the WIFI in their office so that a WIFI-based protocol may be recommended. In this case, a WIFI-based protocol may provide better security and privacy protection for the employees. In contrast, on a bus, BLE-based protocol may be a better choice because there may not be a WIFI access point or the passengers may not trust it even if it exists.

By design, this new venue-based ACT system architecture has the following benefits.

\begin{itemize}
    \item First of all, by allowing every venue to choose the most appropriate location tracking technology, false positive and false negative results could be reduced.

    \item Secondly,  many security attacks (e.g. those based on relaying and replaying messages) described in Section \ref{subsec:security} will be  mitigated because accessibility to users' apps will be restricted. In addition, we can expect the venues to deploy standard cybersecurity countermeasures to further reduce the security risks.

    \item Thirdly, the \emph{persistent surveillance} issue described in Section \ref{subsec:privacy} is naturally eliminated. Furthermore, the \emph{targeted surveillance} issue is also mitigated because it is more difficult to capture messages from targeted users.
\end{itemize}

We acknowledge that this new design will increase the workload and complexity for the users (e.g. they may be required to install different apps from different venues). This can be considered as a tradeoff which is inevitable in order to avoid the identified issues from previous sections.

\subsection{Accountable Mediating}
\label{subsec:mediate}

For various reasons, existing ACT solutions are often implemented as open systems, which means that users can join and quit anonymously at any time. Due to the lack of authentication, messages can be manipulated in various ways without being detected. These features create the following dilemma. On one hand, they protect users' privacy and other fundamental rights because of the anonymity, but on the other hand they drastically expand the attack surfaces against security and privacy. To resolve this dilemma, we propose an \emph{accountable mediating} feature, which requires every venue to act as a mediator to facilitate the contact tracing service. Note that a similar ``mediator" concept can be found in manual contact tracing services. For example, an airliner could record the passengers' contact details for its flights. Later on, if one passenger is tested positive, then other passengers on the same flight will be notified. As another example, it is quite common that users will be required to register their contact information before accessing public facilities such as a zoo or a museum. If so, when one visitor is tested positive then other visitors will be alerted.

With the help of mediators, a venue-based ACT solution exhibits the following valuable capabilities in addition to those mentioned in Section \ref{subsec:trace}.

\begin{itemize}

\item By design, most existing ACT solutions (in particular, BLE-based ones), mainly aim at providing infection risk scores to the users. Some players, e.g. health authority, may be able to infer more information from the collected data, but the inferred information is usually inadequate for them to make informed decisions. In contrast, mediators can obtain users' consent to collect more fine-grained virus spreading information with a venue-based ACT solution. Consented by the users, mediators and other players (e.g. HA) can run a range of secure multi-party computation and/or privacy-preserving data analysis protocols to obtain valuable insights into the desired aspects of virus spreading. By doing so, it overcomes the aforementioned drawbacks and furthermore creates opportunities to solve broader interface and usability issues described in Section \ref{subsec:interface}.

\item If a venue is accountable, the \emph{population coverage} and \emph{transmission path coverage} issues described in Section \ref{subsec:coverage} can be mitigated by integrating a similar manual contact tracing procedure in parallel.  The venue can establish out-of-band communication channels (e.g. phone or email) with its visitors who may or may not have an app installed. If there is an infected user detected on site, the venue can try to contact its visitors (potentially upon their consents). If these visitors are cautious about their health, they may perform a test even if they do not have a high infection risk score or do not have the app installed.

\end{itemize}

One natural concern that getting venues involved in an ACT solution may result in new privacy issues. From a legal perspective, we can require any data collection and usage should be based on users' consent and venues' actions should be audited. In addition, minimisation of information leakage should be taken into account in designing the protocols for the venues (e.g. the BLE-based protocol in next section).

\section{Instantiating the Venue-based ACT Concept}
\label{sec:instantiation}

In this section, we first describe the workflow for our new ACT solution by wrapping up the activities specific to underlying location tracking technology into three general sub-protocols. We then provide an instantiation for these sub-protocols by assuming BLE as the technology in use.

\subsection{The New Workflow}
\label{subsec:newworkflow}

We assume the intended ACT solution will be used in a country or a region and its execution is supervised by a health authority (HA). We further assume two types of technical third parties.  The first one is a back-end server, which is dedicated to managing the aggregated risk information from all the venues. The second one is test centers, which can test users and certify the fact of virus infection. When HA plans the venue-based ACT solution, it first identifies the types of venues which should deploy it, then selects the appropriate location tracking technologies (e.g. WIFI or BLE) for these venues, designs the technology-specific protocols and finally deploys them in the venues respectively.

The technology-specific protocols can differ a lot in their technical details, e.g. how physical distances among users are calculated and what kind of data will be generated and stored. Nevertheless, we expect all these protocols to follow a similar four-phase workflow to that described in Section \ref{subsec:arch}.  Note that we wrap the technology-specific parts into generic sub-protocols, namely $\mathsf{Sense}_v$, $\mathsf{Report}_v$ and $\mathsf{Trace}_v$. We have used a subscript $v$ to denote the protocol for venue $v$, for the purpose of illustrating the fact that protocols differ depending on the underlying location tracking technologies the venues would support.  However, if two venues adopt the same technology (e.g. BLE), then they will deploy the same protocol.

\begin{itemize}
\item In the \emph{initialisation} phase,  HA generates a key pair $(PK_h, SK_h)$ for a signature scheme $(\mathsf{Keygen}, \mathsf{Sign}, \mathsf{Verify})$ and acts as the root of trust. Every venue $v$ generates its own key pair $(PK_v, SK_v)$ for the same signature scheme and gets $PK_v$ certified by HA. Similarly, every test center $t$ generates its own key pair $(PK_t, SK_t)$ for the same signature scheme and gets $PK_t$ certified by HA. As a standard practice, we assume that every venue or test center has a unique identifier that is also certified together with its signing key. In addition, HA generates parameters for a commitment scheme $(\mathsf{Commit}, \mathsf{Reveal})$, e.g. for the Pedersen scheme \cite{Pedersen91}. Every user, say user $i$, generates a commitment as her random identifier $rid_i=\mathsf{Commit}(id_i)$ where $id_i$ is the true identifier say social security number.


\item In the \emph{sensing} phase, when a user $i$ enters into the physical premise of venue $v$, she can optionally give her consent to participate in the sub-protocol $\mathsf{Sense}_v$ which is executed among herself, other users who have also consented, and the venue $v$.  The aim of this sub-protocol is to generate and maintain the contact information of all the users who have visited the venue, and to further facilitate the operations in other phases. Note that venue $v$ is responsible for executing the sub-protocol $\mathsf{Sense}_v$, and the protocol execution can last for a long period of time (e.g. until the end of the pandemic). In contrast, user $i$'s participation in the protocol will only last for her stay in venue $v$ (i.e. her participation stops as soon as she leaves the venue).


\item In the \emph{reporting} phase, suppose that user $i$ has been tested positive by a test center $t$, then they take the following steps to generate a signature if the test result is positive.
\begin{enumerate}
    \item User $i$ sends $rid_i$ to the test center $t$ and proves that it is a commitment of her true identifier $id_i$.
    \item If the proof passes, the test center $t$ generates the following signature for user $i$.
        \begin{displaymath}
        \mathsf{Sign}(contagious\mbox{-}period||rid_i; SK_t),
        \end{displaymath}
    where the string $contagious\mbox{-}period$ stands for the time interval when user $i$ is contagious.
\end{enumerate}

For every venue $v$ that she has visited during the contagious period, user $i$ executes a sub-protocol $\mathsf{Report}_v$ with the back-end server to report her test result.

\vspace{0.1cm}

\item In the \emph{tracing} phase, after receiving a report from any user $i$, the back-end server executes a sub-protocol $\mathsf{Trace}_v$ to inform the at-risk users about their risks. Besides, under the coordination of HA, the back-end server and selected venues can run a range of privacy-preserving data analysis protocols to gain more insights into the pandemic.
\end{itemize}

The main difference between this new workflow and that described in Section \ref{subsec:arch} is that we have required every user to commit to her true identifier (e.g. $id_i$) and use the commitment (e.g. $rid_i$) as her actual identifier in the \emph{sensing}, \emph{reporting} and \emph{tracing}  phases. To certify users' test results, we require users to prove the relationship between their true identifiers and actual identifiers.  Another difference is that venue $v$ may collect more information from its users and engage in secure protocols to collaboratively generate valuable insights in the \emph{tracing} phase. This is not possible in the existing ACT solutions.

\subsection{BLE-based Protocol Design}
\label{subsec:protocol}

We instantiate the $\mathsf{Sense}_v$, $\mathsf{Report}_v$ and $\mathsf{Trace}_v$ sub-protocols for venues which use BLE technology to  monitor the contact history of their users. In brief, users exchange BLE messages to enable contact tracing, while they use standard communication technologies to communicate with the venues and the back-end server. These sub-protocols have a similar flavour to the DP-3T solution, while we have integrated authentication and certification mechanisms to improve the security and privacy guarantees.


\subsubsection{$\mathsf{Sense}_v$ Sub-protocol}

Besides the parameters generated in the \emph{initialisation} phase described in Section \ref{subsec:newworkflow}, this sub-protocol needs some additional parameters. Let $\mathsf{H}$, $\mathsf{PRG}$ and $\mathsf{PRF}$ denote a cryptographic hash function, a pseudorandom number generator and a pseudorandom function, respectively. In addition, there is an epoc parameter $L$ and a time window parameter $W$. During each epoc the same pseudorandom identifier will be broadcast repeatedly, while in each time window a new private key is generated to construct the pseudorandom identifiers in its epocs. For example $L$ and $W$ can be set to be 3 minutes and 120 minutes respectively.

The protocol is initiated by venue $v$, which deploys standard Bluetooth devices to capture all the BLE messages on its physical premise. Venue $v$ periodically broadcasts its certificate so that the users can use it when they leave the venue.  During the protocol execution, users and venue $v$ perform as follows.  Note that we use user $i$ as an example for the description while other users perform in the same manner.
\begin{itemize}

    \item When user $i$ enters into venue $v$, she gives consent to the ACT app on her mobile device to broadcast and receive BLE messages.   The app generates a private key $SK_{i,v, 1}$ for the first time window and constructs $n=\frac{W}{L}$ pseudorandom identifiers.
    \begin{small}
        \begin{eqnarray}
        \label{ephemeralid}
        EphID_{i, v, 1, 1}||\cdots||EphID_{i, v, 1, n}= \mathsf{PRG}(\mathsf{PRF}(SK_{i,v, 1}, ``broadcast key||id_v"))
        \end{eqnarray}
        \end{small}
    In the $t$-th epoc of the first time window, the app broadcasts $EphID_{i, v, 1, t}$ and collects  pseudorandom identifiers from other users. At the end of the $t$-th epoc, the following identifier record is formed and stored.
    \begin{displaymath}
    (1, t, EphID_{i, v, 1, t}, \mathbb{S}_{1,t}, SK_{i,v, 1}),
    \end{displaymath}
    where the set $\mathbb{S}_{1,t}$ contains all the received identifiers. If user $i$ stays longer than a time window, then her app will generate a new private key for the second time window, construct new pseudorandom identifiers, and broadcast them. This process continues until user $i$ decides to leave venue $v$.

    Note that, after receiving a BLE message, user $i$'s app also stores auxiliary data together with the message. Such data can include signal strength and so on, as in existing BLE-based ACT solutions. We skip the details here.


    \item On leaving, user $i$ interacts with venue $v$ as follows. We assume the communication is through a unilaterally secure channel, which guarantees that exchanged messages are protected with confidentiality and integrity while user $i$ can stay anonymous.

    \begin{enumerate}
        \item User $i$'s app sends the following record to venue $v$ to notify her leave, where $nonce_i$ is computed as a commitment of $rid_i$ and  $time_i$ is the current time (i.e. $nonce_i=\mathsf{Commit}(rid_i)$).
        \begin{displaymath}
        (nonce_i, time_i, \mathsf{H}(EphID_{i, v, 1, 1}||\cdots||EphID_{i, v, x, y}))
        \end{displaymath}

        The app locally stores $(nonce_i, time_i, id_v)$ together with the identifier records from all eopcs, as well as other auxiliary information.


        \item Venue $v$ generates and sends the following signature to user $i$.
        \begin{eqnarray}
            \label{presencesign}
        \mathsf{Sign}(nonce_i||time_i||\mathsf{H}(EphID_{i, v, 1, 1}||\cdots||EphID_{i, v, x, y}); SK_v)
        \end{eqnarray}


        \item After receiving the confirmation signature, user $i$'s app stores it locally and then halts so that no messages will be broadcast or received anymore.
    \end{enumerate}


    \item Venue $v$ sets a retention threshold for the received pseudorandom identifiers, e.g. 14 days. Periodically (e.g. daily), venue $v$ encodes all pseudorandom identifiers from its database into a bloom filter $\mathbb{BF}_{v}$ and sends it to HA. 
\end{itemize}

\subsubsection{$\mathsf{Report}_v$ Sub-protocol}

Referring to Section \ref{subsec:related}, we adopt a \emph{upload-what-you-sent} approach.
When an infected user $i$ wants to report her pseudorandom identifiers from the contagious period to the back-end server, she follows the steps below.
\begin{enumerate}
    \item For each venue $v$ she has visited during the contagious period, user $i$ sends the following information to the backend-server.
    \begin{displaymath}
    (contagious\mbox{-}period||rid_i, \mathsf{Sign}(contagious\mbox{-}period||rid_i; SK_t)),
    \end{displaymath}
    \begin{displaymath}
    (nonce_i, \mathsf{Reveal}(nonce_i)),
    \end{displaymath}
    \begin{displaymath}
    (time_i, \mathsf{Sign}(nonce_i||time_i||\mathsf{H}(EphID_{i, v, 1, 1}||\cdots||EphID_{i, v, x, y}); SK_v)),
    \end{displaymath}
    \begin{displaymath}
    (id_v, y, SK_{i,v, 1}, SK_{i,v, 2}, \cdots, SK_{i,v, x}),
    \end{displaymath}
    where $id_v$ indicates that this record is linked to venue $v$ and $y$ indicates the number of epocs in the last time window.


    \item After receiving the information from user $i$, the back-end server does the following.
    \begin{enumerate}
        \item Verify the signature $\mathsf{Sign}(contagious\mbox{-}period||rid_i; SK_t)$, and verify that $nonce_i$ is a commitment of $rid_i$ based on $\mathsf{Reveal}(nonce_i)$. If any verification fails, halt.

        \item Construct the pseudorandom identifiers based on $(SK_{i,v, 1}, \cdots, SK_{i,v, x})$ and verify  $\mathsf{Sign}(nonce_i||time||\mathsf{H}(EphID_{i, v, 1, 1}||\cdots||EphID_{i, v, x, y}); SK_v)$. If the verification fails, halt.

        \item Run a two-party matching protocol with HA to assure that the identifiers in the previous step have indeed appeared in venue $v$, i.e. these identifiers have been encoded into the bloom filter $\mathbb{BF}_{v}$.


        \item If the above check succeeds, the back-end server adds the following record to its database.
            \begin{eqnarray}
            \label{backendrecord}
            (id_v, time_i; EphID_{i, v, 1, 1}, \cdots, EphID_{i, v, x, y})
            \end{eqnarray}

        \item The back-end server notifies venue $v$ so that it can take the necessary actions to eliminate any residual risks (e.g. clean its environment and encourage its staff to perform a test).
    \end{enumerate}

\end{enumerate}

\subsubsection{$\mathsf{Trace}_v$ Sub-protocol}

Periodically (e.g. daily),from the back-end server, user $j$ retrieves information about the infected users for each venue $v$ she has visited.
\begin{enumerate}
    \item In order to prove his presence on the physical premise of venue $v$ at time $time_j$, user $j$ sends $(nonce_j, time_j, \mathsf{H}(EphID_{j, v, 1, 1}||\cdots), id_v)$ and the signature $\mathsf{Sign}(nonce_j||time_j||\mathsf{H}(EphID_{j, v, 1, 1}||\cdots); SK_v))$ to the back-end server.

    \item The back-end server verifies $\mathsf{Sign}(nonce_j||time_j||\mathsf{H}(EphID_{j, v, 1, 1}||\cdots); SK_v)$. If the verification fails, halt.


    \item The back-end server checks every record (which is in the form of (\ref{backendrecord})), whose first element is $id_v$, and returns the pseudorandom identifiers to user $j$ if the record's second element (i.e. the timestamp $time_i$) satisfies certain condition with respect to $time_j$. Note that the condition can depend on the characteristics of venue $v$. For instance, if venue is a supermarket, the condition can be $time_j$ and the record's timestamp have the same date.

    \item With all the retrieved pseudorandom identifiers, user $j$ locally evaluates his infection risk at the venue $v$, and acts accordingly.
\end{enumerate}

\subsection{Brief Security and Privacy Analysis}

Our analysis in Section \ref{sec:funchallenges} and Section \ref{sec:spchallenges} indicates a number of trade-offs, e.g. security-privacy, interface-privacy, usability-security/privacy, and complexity--security/privacy. Motivated by these facts, the proposed sub-protocols neither aim at eliminating all the issues nor intend to focus on one specific set of issues. Instead, we make use of simple cryptographic primitives such as signature and commitment schemes to reduce the attack surface. For clarity, we summarize the players' capabilities in Table \ref{tab:capability}.

\begin{table}[H]
\small
\begin{center}
\begin{tabular}{|l|m{9.2cm}|}\hline
\textbf{Player}   & \textbf{Capability}    \\    \hline
HA &   Certify venues and test centers; Store venues' Bloom filters in the $\mathsf{Sense}_v$ sub-protocol;  Verify the presence of pseudorandom identifiers in the $\mathsf{Report}_v$ sub-protocol;  Coordinate data analysis in the $\mathsf{Trace}_v$ sub-protocol\\    \hline
Back-end server & Receive, verify, and distribute pseudorandom identifiers from infected users in the $\mathsf{Report}_v$ and $\mathsf{Trace}_v$ sub-protocols\\ \hline
Venues & Collect pseudorandom identifiers from their visitors; Aggregate these identifiers into Bloom filters; Certify their visitors' presence; Deploy necessary cybersecurity countermeasures \\ \hline
Honest user & Collect pseudorandom identifiers from her contacts; Retrieve her contacts' pseudorandom identifiers when they are tested positive\\ \hline
Malicious user & Inject arbitrary pseudorandom identifiers (could be detected by accountable venues in the $\mathsf{Sense}_v$ sub-protocol, could be detected by the back-end server in the $\mathsf{Report}_v$ sub-protocol);  Provide fake nonces (could not submit info in the $\mathsf{Report}_v$ sub-protocol)\\ \hline
\end{tabular}
\caption{Capabilities of Different Players}
\label{tab:capability}
\end{center}
\end{table}

Next, we briefly discuss the security and privacy properties of our BLE-based protocol, particularly the sub-protocols from Section \ref{subsec:protocol}.  In our analysis, we make a general assumption that users' communication with venues and back-end server is via anonymous channel, as assumed in the $\mathsf{Sense}_v$ sub-protocol. Particularly, different communication sessions (e.g. those with two venues) cannot be linked. Note that this assumption is widely adopted by most existing ACT solutions. Furthermore, we assume that HA, the back-end server and venues are accountable and semi-honest in the cryptographic sense, namely the venues will not behave maliciously such as colluding with each other or with users.

\subsubsection{Security Analysis}

By putting venues in the center to mediate contact tracing, many security vulnerabilities in existing solutions have been avoided given that the venues act in an accountable manner. At a very high level, a venue can monitor the pattern of messages on its physical premise to detect misbehaving activities, e.g. an attacker broadcasts an enormous amount of messages to mount DDoS attacks. Furthermore, venues can create their own policies and deploy cybersecurity tools to maximize the security protection for their users. For example, a venue can monitor the signal strength of its users' devices and detect those whose signal strength is too strong (e.g. an attacker can use such a device to result in abnormal false positives).  At the protocol level, our solution improves the security protection in the following aspects.
\begin{itemize}
    \item To some extent, the association between users and venue $v$ is certified in the \emph{sensing} phase, i.e. through the signature message (\ref{presencesign}) in the $\mathsf{Sense}_v$ sub-protocol. It is possible for user $i$ to ``impersonate" user $j$ by using $rid_j$ instead of $rid_i$ in the $\mathsf{Sense}_v$ sub-protocol. However, with the help of timestamps, the same user cannot simultaneously appear in two different venues in the following sense. Assume user $j$ is tested positive, he was at venue $v'$ at time $time_j'$ and user $i$ impersonated him at venue $v$ at $time_j$ which is very close to $time_j'$.  Note that this means user $i$ and user $j$ collude. Under this assumption, it is impossible for user $j$ to submit two records for venue $v$ and $v'$ respectively in the $\mathsf{Report}_v$ sub-protocol.  The timestamps in the leaving procedure in the $\mathsf{Sense}_v$ sub-protocol also introduces other timing constraints. For instance, at $time_i$, it is impossible to forge any leaving event with a timestamp $time_i'$ which is earlier than $time_i$ as long as the venues are honest.
    \vspace{0.1cm}

    \item In the construction of the pseudorandom identifiers at venue $v$, its identifier $id_v$ is required to be included in Equation (\ref{ephemeralid}). Therefore, it is not useful for an attacker to relay or replay pseudorandom identifiers from one venue to another, because the back-end server will only accept an infected user's pseudorandom identifiers for venue $v$ if they are generated based on $id_v$ in the $\mathsf{Report}_v$ sub-protocol. If the attacker does so, the only harm it can result in is making venue $v$ store some useless information in its bloom filter $\mathbb{BF}_{v}$.

    \vspace{0.1cm}

    \item When an infected user $i$ submits her pseudorandom identifiers to the back-end server in the $\mathsf{Report}_v$ sub-protocol, she needs to present the ephemeral keys. Based on the security properties of pseudorandom number generator  $\mathsf{PRG}$ and pseudorandom function $\mathsf{PRF}$, she cannot submit messages from other users without knowing the ephemeral keys behind. Furthermore, the back-end server only accepts pseudorandom identifiers that have already collected by a venue. Therefore, an infected user can not submit arbitrary pseudorandom identifiers of her choice.

\end{itemize}

It is clear that our protocol cannot cryptographically binding human users, mobile devices and pseudorandom identifiers they broadcast. This means it is still vulnerable to sophisticated attacks. For instance, user $i$ can share her credential with user $j$ who can then ``impersonate" her in different venues. Such attacks cannot be detected as long as user $i$ does not submit records with very close timestamps as we have shown above. Such security vulnerabilities could be resolved if we introduce trusted hardware and other techniques into the solution, however this will degrade the usability and coverage issues and may cause new privacy concerns.

\subsubsection{Privacy Analysis}

By design, the \emph{data minimisation} principle has been taken into account because a user can only retrieve information about the infected users if they have visited the same venues. As a remark, asking venue $v$ to record BLE messages on its physical premise does not introduce new privacy risks because it can does the same in any existing BLE-based solutions. However, the example concern regarding this principle described in Section \ref{subsec:privacy} may still exist. To resolve the issue, we can require user $i$ to report both her arrival and departure timestamps in the $\mathsf{Sense}_v$ sub-protocol, instead of only departure timestamp $time_i$. As a result, the time interval can be used by the back-end server to further validate the submitted identifiers in the $\mathsf{Report}_v$ sub-protocol, and the back-end server can control user $j$'s request in the $\mathsf{Trace}_v$ sub-protocol (e.g. if his stay at venue $v$ is too short then no information will be retrieved). Unfortunately, this enhancement may introduce new privacy concerns because it faithfully describes every user stay at every venue.  We leave a further investigation of this matter as a future work. Regarding the \emph{data minimisation} principle, there is a possibility that colluded users can aggregate the information they have retrieved. We can foresee two extensions to overcome this vulnerability and rigorously enforce the \emph{data minimisation} principle. One is to employ trusted hardware so that users cannot aggregate their information at their will. This approach has usability problem as we mentioned before. The other is to let users run a secure two-party computation protocol with the back-end server to evaluate their infection risks. With this approach, the computational complexity for the back-end server will be an obstacle if many users query their results simultaneously. To our knowledge, it remains as an open problem to craft a usable and efficient solution for this vulnerability.

In the following, we elaborate on other privacy enhancements our solution can provide.

\begin{itemize}
    \item With respect to an uninfected user $j$, as a result of the \emph{hiding} property of the commitment scheme, the $nonce_j$ submitted to venue $v$ leaks no information in the $\mathsf{Sense}_v$ sub-protocol because it is a random number from the perspective of the venue. Similarly, there is no leakage of identifier information when the user retrieves information from the back-end server in the $\mathsf{Trace}_v$ sub-protocol. In comparison to existing solutions,  user $j$ has less exposure to the potential attackers because his app only broadcasts and receives messages in selected venues. Therefore, the \emph{ephemeral linkage} issue from Section \ref{subsec:privacy} is partially mitigated since there is no link between a user's messages in any two venues.

   \vspace{0.1cm}

    \item With respect to an infected user $i$, the situation is a bit more complex. Under our assumption, the back-end server does not learn any identifier information about user $i$ in the $\mathsf{Report}_v$ sub-protocol because both $nonce_i$ and $rid_i$ look like random number except the fact $nonce_i$ is a commitment of $rid_i$. Comparing to existing solutions, user $i$ faces much less privacy risks because only those users who have been in the same venues during her contagious period can potentially figure out her infection status. Now, suppose the back-end server and test centers collude, then the fact that user $i$ has been infected and also submitted her pseudorandom identifiers is known to the colluded parties. In our opinion, this might not be a privacy issue.
    \begin{itemize}
        \item If there is a rule saying that the user must share her pseudorandom identifiers if she is tested positive, then this is not a privacy leakage. Instead, it is a necessary measure to enforce that user $i$ indeed submits her information. It also makes users act accountably.

        \item If user $i$ can decide whether or not to submit her information, then leaking the fact may not cause any harm to her as it has been an option for her. Nevertheless, cryptographic techniques can be applied to hide this fact and we leave it as a future work.
    \end{itemize}

    \vspace{0.1cm}

    \item With respect to venue $v$, the following information is revealed.
    \begin{itemize}

        \item If it is optional for venue $v$ to decide whether or not to participate in the solution. One privacy concern is that HA learns venue $v$'s option because it should periodically submit  $\mathbb{BF}_{v}$ in the $\mathsf{Sense}_v$ sub-protocol if it is participating. In practice, if venue $v$ is accountable, then it should participate to protect both itself and its visitors. In particular, a venue should be informed whether some recent visitor has been tested positive, sho that it could take more actions to sanitize its environment to block all potential transmission paths.  We believe that it should be mandatory for public facilities to participate in the solution.
        \item If venue $v$ participates in the ACT solution, the back-end server and HA learn the fact whether any visitor of venue $v$ has been tested positive in the $\mathsf{Report}_v$ sub-protocol. If venue $v$ is accountable (or, even implied by the law), then this should not be considered as a privacy concern. In practice, such information will be revealed anyhow. If user $i$ is tested positive, then some human tracing experts will ask her about the venues she has visited and then take more actions accordingly. Therefore, this is not a privacy concern in practice.

        \item If venue $v$ has a visitor, say user $i$, tested positive, then this fact may be revealed to other visitors who have visited it in the same period. The revelation will only happen if user $i$ reports his infection to the back-end server. Technically, this kind of information can also be inferred in most existing ACT solutions if the users match their local information with that from the third parties (e.g. the back-end server in DP-3T).  Similar to the above case, this may not a privacy concern in practice.

    \end{itemize}

     We note that it is possible to employ cryptographic techniques to hide the more information from HA and the back-end server. As always, there will be a trade-off with efficiency of the solution.

\end{itemize}

\section{Conclusion}
\label{sec:con}

In this paper, we first reviewed the existing ACT concept and some recent solutions, and then elaborated on the functional as well as security and privacy issues these solutions face in practice. We went further to propose a venue-based concept, which mitigates some of the identifier issues by design and also introduces the concept of accountability for venues to actively participate in the solution. We finally instantiated the new venue-based ACT concept by providing a BLE-based protocol. Based on our analysis, we believe that a venue-based ACT solution can partially address the functional issues and also provide higher level of security and privacy guarantees in comparison to existing ACT solutions. Nevertheless, we acknowledge that many improvements can be done with existing technologies, e.g. introducing trusted hardware and integrating more advanced cryptographic techniques. It is an interesting direction for scientific research. On the other hand, there are various conflicting points among different requirements, e.g. the functional ones and security/privacy ones, and between security and privacy. It is another interesting research direction to investigate the possible tradeoffs among all the requirements and study users' acceptance attitude towards various design choices. By getting venues involved, the venue-based ACT concept creates other interesting opportunities. For example, instead of requiring its visitors to install an app on their own devices, a venue can offer some mobile devices (e.g. UWB bracelet) to its visitors so that they can record their contact history on its premise. This is another interesting direction for research and development.

\section*{Acknowledgement}
This work is partially funded by the European Unions Horizon 2020 SPARTA project, under grant agreement No 830892.

\bibliographystyle{plain}
\bibliography{act}

\end{document}